\documentclass[12pt]{iopart}

\usepackage{xcolor,graphicx}
\usepackage{amssymb,bbm}
\usepackage[utf8]{inputenc}
\usepackage{comment}
\usepackage{braket}
\usepackage[normalem]{ulem}
\usepackage[tableposition=top]{caption}
\usepackage{acronym}
\usepackage{amsfonts}
\usepackage{amsthm}
\usepackage{caption}
\usepackage{subcaption}
\usepackage[colorlinks=true,linkcolor=blue,urlcolor=blue,citecolor=blue,pdfusetitle]{hyperref}
\usepackage{cleveref}

\crefname{equation}{Eq.}{Eqs.}
\crefname{figure}{Fig.}{Figs.}

\definecolor{sm}{rgb}{0.0, 0.0, 1.0}

\acrodef{mlp}[MLP]{Multi Layer Perceptron}
\acrodef{cnn}[CNN]{Convolutional Neural Network}
\acrodef{mse}[MSE]{Mean Squared Error}
\acrodef{mae}[MAE]{Mean Absolute Error}
\acrodef{mre}[MRE]{Mean Relative Error}
\acrodef{mlp}[MLP]{Multilayer Perceptron}

\begin{document}

\title{Deep learning enhanced noise spectroscopy of a spin qubit environment} 

\author{Stefano Martina$^{1,2,*}$, Santiago Hern\'{a}ndez-G\'{o}mez$^{3,2,*}$, Stefano Gherardini$^{4,2}$, Filippo Caruso$^{1,2}$, Nicole Fabbri$^{5,2}$}
\address{$^1$ Dipartimento di Fisica e Astronomia, Universit\`a di Firenze, I-50019, Sesto Fiorentino, Italy}
\address{$^2$ European Laboratory for Non-linear Spectroscopy (LENS), Universit\`a di Firenze, I-50019 Sesto Fiorentino, Italy}
\address{$^3$ Research Laboratory of Electronics, Massachusetts Institute of Technology, Cambridge, MA 02139}
\address{$^4$ Istituto Nazionale di Ottica del Consiglio Nazionale delle Ricerche (CNR-INO), Area Science Park, Basovizza, I-34149 Trieste, Italy}
\address{$^5$ Istituto Nazionale di Ottica del Consiglio Nazionale delle Ricerche (CNR-INO), I-50019 Sesto Fiorentino, Italy}
\address{$^*$ Equally contributed to this work}
\eads{\mailto{stefano.martina@unifi.it}, \mailto{shergom@mit.edu}, \mailto{stefano.gherardini@ino.cnr.it}, \mailto{filippo.caruso@unifi.it}, \mailto{fabbri@lens.unifi.it}}

\vspace{10pt}
\begin{indented}
\item[]\today
\end{indented}

\begin{abstract}
The undesired interaction of a quantum system with its environment generally leads to a coherence decay of superposition states in time. A precise knowledge of the spectral content of the noise induced by the environment is crucial to protect qubit coherence and optimize its employment in quantum device applications.
We experimentally show that the use of neural networks can highly increase the accuracy of noise spectroscopy, by reconstructing the power spectral density that characterizes an ensemble of carbon impurities around a nitrogen-vacancy (NV) center in diamond. Neural networks are trained over spin coherence functions of the NV center subjected to different Carr-Purcell sequences, typically used for dynamical decoupling (DD). As a result, we determine that deep learning models can be more accurate than standard DD noise-spectroscopy techniques, by requiring at the same time a much smaller number of DD sequences.  
\end{abstract}

\vspace{2pc}

%\maketitle

\section{Introduction}

Quantum sensing combines theoretical results with experimental and engineering techniques to carry out inference of signals with improved accuracy and/or less computation time by making use of quantum physics~\cite{DegenRMP2017,GomezFrontiers2021}. 

A quantum sensor takes advantage of the fragility of its quantum properties, such as quantum coherence or entanglement, to improve the detection of external perturbations with higher accuracy compared to any classic sensor. 
However, this same property implies that the quantum sensor is subjected to detrimental noise stemming from the coupling with its environment. 
For this reason, it is desirable to fully characterize the sensor's environment, either to filter out its detrimental effect, or to take it into account when detecting external signals, for example, in algorithms using quantum optimal control~\cite{PoggialiPRX2018,MuellerSciRep2018,RemboldAVSQ2020,MarshallArXiv2021,ZhangAAPPS2023}.  

Neural networks (NNs)~\cite{BishopPRML2006,GoodfellowDL2016}, i.e., algorithmic models provided by the interconnection of a group of nodes commonly called neurons, could be a powerful tool to infer the sensor's environment. In this context, deep learning has been already proposed theoretically for the classification and detection of quantum noise features~\cite{YoussryNpjQI2020,MartinaArXiv2021,WisePRXQ2021}, and employed experimentally for the following tasks. \textit{(a)} Estimating the spectra of minuscule amounts of complex molecules~\cite{AharonSciRep2019} for nano nuclear magnetic resonance; \textit{(b)} the sensing of magnetic-field strength at room temperature with high precision~\cite{SantagatiPRX2019,Jung21} by using nitrogen vacancy (NV) centers; \textit{(c)} performing error mitigation~\cite{StrikisarXiv2021} and noise learning~\cite{HarperNatPhys2020,MartinaQMI2022_learning,MartinaArXiv2021_2}; \textit{(d)} the tracking of quantum trajectories~\cite{KoolstraarXiv2021}; \textit{(e)} classification of many-body quantum states~\cite{GongArXiv2022} in superconducting quantum circuits; \textit{(f)} improving quantum error correction~\cite{DingQE2022}. Also quantum neural networks are recently investigated in order to solve a given quantum technology task with a greater accuracy than classical NNs~\cite{TilayeQE2023,ZhangQE2021,WeiAAPPS2022}. However, to our knowledge, experimental noise spectroscopy in single color centers in diamond via deep learning is still missing.

\begin{figure}
    \centering
    \includegraphics[width=1.0\textwidth]{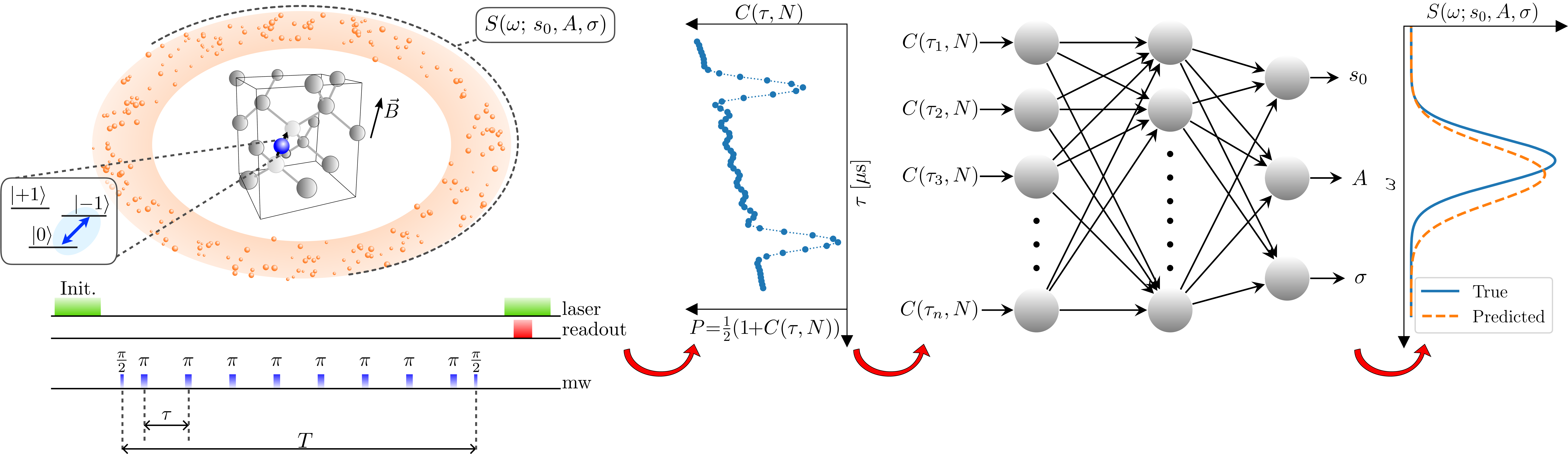}
    \caption{\textbf{NV center and Neural Networks for noise spectroscopy.} 
    % in the fig: C(\tau_i,N). In caption: here we show only the case for a fixed value of N, changing \tau. Later we show that we consider also different values of N.
    The NV center is surrounded by an ensemble of $^{13}C$ nuclear spins (orange spheres) that collectively induce dephasing to the NV electronic spin (blue sphere). The NV electronic spin is controlled with a DD sequence (specifically, a Carr-Purcell (CP) sequence) with the aim to measure its dephasing, and therefore characterize the NSD of the nuclear spin bath, i.e., $S(\omega; s_0,A, \sigma)$. The CP sequence is formed by $N$ equidistant $\pi$ pulses in between an initial and a final $\pi/2$ pulse. The time $\tau$ between the $\pi$ pulses determines the measurement total time $T=N\tau$, given that the time between the first $\pi/2$ and the train of $\pi$ pulse and the time between the last $\pi$ and $\pi/2$ pulses are both equal to $\tau/2$. 
    Then, we measure the output of this experiment, which is the probability $P = \frac{1}{2}(1+C(\tau,N))$ that the NV center remains in the initial state $\ket{0}$. %The spin coherence function $C(t)$ -- evaluated at previously-determined times in the set $T\in\{t_1,t_2,\ldots,t_n\}$ (the $t_k$'s are obtained by changing $\tau$ with $N$ fixed)
    The spin coherence function $C(\tau,N)$ -- evaluated at previously-determined inter-pulse times in the set $\tau\in\{\tau_1,\tau_2,\ldots,\tau_n\}$ is the input of the designed Neural Networks (NNs). For illustrative purposes, here we only consider one fixed value of $N$. In our study we also consider a set of different values of N [Sec.~\ref{subsec:training}].} 
    After being trained, the NNs return the estimation of the NSD parameters.
    \label{fig:fig1}
\end{figure}

In this paper, we demonstrate that NNs can be used to process the data obtained by a qubit, operating as a quantum sensor, and then reconstruct the noise spectrum that induces dephasing into the qubit itself. 
In particular, we focus on a qubit under dynamical decoupling (DD) control sequences~\cite{ViolaPRA1998,FaoroPRL2004} in the presence of classical random noise with an unknown power density spectrum, usually denoted as \emph{noise spectral density} (NSD). 
Beyond testing numerically our machine learning models, we use a single NV center in diamond as a spin qubit sensor and we perform a spectroscopic reconstruction of the magnetic noise of its local environment. The latter comprises $^{13}$C nuclear spins randomly distributed in the diamond lattice~\cite{TaylorNatPhys2008,GoldsteinPRL2011,AbobeihNatComm2018} (see \cref{fig:fig1}). The dephasing affecting the qubit sensor is analyzed by applying a set of DD control pulses that realize filter functions~\cite{ViolaPRA1998,FaoroPRL2004,CywinskiPRB2008,BiercukJPB2011} in the frequency domain. The filter functions are designed to select specific noise components, without sensing all other system-bath interactions. A widely used DD control pulse is the Carr-Purcell (CP) sequence~\cite{CarrPR1954,DegenRMP2017} that is given by $N$ equidistant $\pi$ pulses, performed between an initial and a final $\pi/2$ pulse. CP sequences act in the frequency domain approximately as Dirac comb filters~\cite{MazeNature2008}; hence, they have been used to perform spectroscopy of intricate signals, e.g., for noise spectroscopy~\cite{YugePRL2011,AlvarezPRL2011}. With this protocol, the requirement to achieve high values of the noise reconstruction accuracy is to perform sequences with a high number of pulses meaning $N \in [30, 120]$ (as in Ref.~\cite{HernandezPRB2018}) or higher, so that the Dirac comb filter approximation remains valid (in fact, $N$ determines the filter width). 
This usually leads to long experiments to reconstruct the whole spectrum of the noise. Other techniques using non-equidistant or even more sophisticated DD sequences~\cite{MuellerSciRep2018,DoNJP2019,FreyNatComm2017,FreyPRApp2020,WangArXiv2022} have proved to be effective for noise sensing, but sometimes at the price of a higher computational burden.  

For our sensing task, NNs are designed to solve a regression problem, i.e., the reconstruction of the NSD. Here, we assume that the NSD of the bath of spins has a Gaussian profile~\cite{HernandezPRB2018,Szankowski17,Szankowski18}. The Gaussian NSD is thus parametrized as a function of key parameters, i.e., the mean value, variance, offset and noise power that we aim to reconstruct. 
Note that our proposal can be adapted to other parametrized NSD functions. The NNs are trained over a set of synthetic data generated by simulating how the coherence of the qubit sensor decays over time under the influence of both the CP control pulses and the NSD. Moreover, to make the measurement statistics as close as possible to the ones obtained from the experiments, extra artificial errors sampled from a normal distribution are added. 

Our approach using NNs entails the following advantages that we have proven experimentally. \textit{(i)} NNs have the capability to predict never-before-seen experimental data, and they can work with a better reconstruction accuracy (even up to $7$ times better, as shown in the section Results below) than standard noise spectroscopy, as the \'Alvarez-Suter method~\cite{AlvarezPRL2011}, by making use at the same time of DD control sequence with a much smaller number of pulses. \textit{(ii)} The training dataset, which can contain both synthetic and experimental data, is generated just once and then it can be applied several times, as long as the new collected data reproduce the physical context under analysis. In connection with \textit{(i)}, we are going to show that the amount of data used as input to the NNs can be smaller than the one needed to resolve the NSD by means of standard noise spectroscopy methods.

From our knowledge, this work is the first experimental proof of enhanced reconstruction performance with NNs for carrying out noise spectroscopy in single color centers in diamond. We thus expect that the techniques discussed here could fast become a novel standard spectroscopy tool both for such quantum systems and other quantum platforms in which regression problems have to be solved. 

%%%%%%%%%%%%%%%%%%%%%%%%%%%%%%%%%%%%%%%%%%%%
\section{Results}

\subsection{Generation of training dataset}\label{sec:datasetGeneration}

The training dataset is composed of synthetic data that are originated by simulating the coherence decay of the qubit sensor in a noise spectroscopy experiment based on DD, as the one depicted in Fig~\ref{fig:fig1}. This standard sensing procedure, which stems from Ramsey interferometry~\cite{DegenRMP2017}, maps information about the quantum coherence of the sensor into the population in $|0\rangle$ that is then effectively recorded. 
After having initialized the qubit sensor in the ground state $|0\rangle$, a $\pi/2$ pulse is applied such that the qubit state $|\psi\rangle$ is the superposition $(|0\rangle + |1\rangle)/\sqrt{2}$. Then, we perform a CP control sequence consisting in a train of $\pi$ pulses that flips repeatedly the qubit, and finally, a second $\pi/2$ pulse is applied in order to map the phase of the qubit into its population. The probability that the state of the quantum sensor is $|0\rangle$, which corresponds to the observable population, equals to~\cite{DegenRMP2017,HernandezPRB2018}
\begin{equation}
    P = \frac{1}{2}\left(1 + C(\tau,N)\right),
\end{equation}
where $N$ is the number of $\pi$ pulses and $\tau$ is the time between them. The coherence function $C(\tau,N)$ is simulated numerically, for a set of different values of $\tau$ and $N$, to generate the training dataset.

Let us now introduce the decoherence function that quantifies how the quantum coherence $C(\tau,N)$ is modified under the action of both the external bath of spins and a set of CP control pulses. The control sequence has the effect to modulate the coherence content of the qubit sensor, while the interaction with the bath, associated to the NSD $S(\omega)$, tends on average to destroy such coherence. Overall, under the joint presence of control fields and a noise source, the coherence decays as $C(\tau,N) \equiv e^{-\chi(\tau,N)}$, where $\chi(\tau,N)$ denotes the \emph{decoherence function}~\cite{GordonJPB2007,GordonPRL2008,BiercukJPB2011,DallaPozzaIJQI2019}:
\begin{equation}\label{eq:decoh_function}
\chi(\tau,N) = \int \frac{d\omega}{\pi\omega^2}F(\omega,\tau,N)S(\omega)\,.
\end{equation}
In \cref{eq:decoh_function}, the \emph{filter function} $F(\omega,\tau,N) \equiv |Y(\omega,\tau,N)|^{2}$ is the square modulus of the Fourier transform of the so-called modulation function $y(t,\tau,N)$. The latter is constant piecewise, with values $\pm 1$, and switches sign at the times $t=\tau/2,3\tau/2,\dots,(N-1/2)\tau$ where each $\pi$ pulse is applied~\cite{GomezFrontiers2021}. 
Notice that we are assuming that the $\pi$ pulses are instantaneous, a reasonable assumption for our experimental setup where a $\pi$ pulse duration is $\sim 0.1$~$\mu$s and the time between pulses is $\tau\in [3.3,6.1]$~$\mu$s. 
Let us now recall the expression, in the frequency domain, of the filter function for a CP sequence with even $N$:
\begin{equation}
    %F(\omega) = 8\sin^{2}\left(\frac{\omega T}{2}\right)\sec^{2}\left(\frac{\omega T}{2N}\right)\sin^{4}\left(\frac{\omega T}{4N}\right),
    %
    F(\omega, \tau, N) = 8\sin^{2}\left(\frac{\omega \tau N}{2}\right)\sec^{2}\left(\frac{\omega \tau }{2}\right)\sin^{4}\left(\frac{\omega \tau }{4}\right),
\end{equation}
while for odd $N$, $\sin^{2}(\omega \tau N/2)$ has to be replaced with $\cos^{2}(\omega \tau N/2)$~\cite{CywinskiPRB2008,GomezFrontiers2021}.

In order to generate the training dataset, the NSD $S(\omega)$ is parameterized as
\begin{equation}
    S(\omega) = s_{0} + A\exp\left(-\frac{(\omega - \omega_c)^2}{2\sigma^2}\right).
\end{equation}
Thus, being a Gaussian distribution, the NSD is fully described by the \emph{offset} $s_0$, \emph{amplitude} $A$, \emph{width} $\sigma$ and \emph{center} $\omega_c$. For the training dataset in the paper, the values of these parameters are taken from the following intervals: $s_0\in[4\cdot 10^{-4}, 4\cdot 10^{-3}]$~MHz; $A\in[0.3, 0.7]$~MHz; $\sigma \in [2\cdot 10^{-3}, 9\cdot 10^{-3}]$~MHz. 
%
%; $B\in[403, 403.4]$~G. 
%
Instead, $\omega_c$ is kept constant. This is because in our experimental setup the NSD stems from the interaction with a large ensemble of unresolved $^{13}$C impurities (nuclear spin bath) around the NV electronic spin. Therefore, the center of the NSD corresponds to the Larmor frequency $\omega_c = \gamma B$, where $\gamma = 1.0705$~kHz/G is the gyromagnetic ratio of the $^{13}$C nuclear spins, and $B$ is the amplitude of a static magnetic field aligned with the NV quantization axis, $z$. Such static magnetic field is well known during the experimental procedure since it determines the NV electronic spin resonances ($B=403.2\pm2$ G).

The training dataset is generated by uniformly sampling $10^4$ sets of parameters within the chosen intervals. Hence, overall we consider $10^4$ distinct sequences of NSD parameters that are used to simulate different coherence curves $C(\tau,N)$. These sequences are taken in the time intervals $\tau \in [3.3 , 3.66]~\mu$s and $[5.5 , 6.1]~\mu$s with sampling time $\Delta \tau = 1$~ns ($\Delta \tau = 20$~ns in the experimental case, see below), and for $N=\{1,8,16,24,32,40,48\}$. These intervals are significant for our study because they include the values of $\tau$ at which the coherence decay curve exhibits the first and second order collapses induced on the qubit sensor by the bath of $^{13}$C impurities (for the coherence curves, the first and second order of the collapses refer to the harmonics of the filter functions $F(\omega, \tau, N)$, for more details see Ref.~\cite{HernandezPRB2018}). 
Finally, in order to make the synthetic data used to train the NNs closer to the experimental setting, extra 
artificial errors sampled from a normal distribution with zero expected value and standard deviation equal to $0.05$ (comparable with the expected error in our experimental measurements) are added to every point of the generated coherence decay curves. In this way, one may mitigate the over-fitting of the employed machine learning models that are thus expected to better generalize to unseen data. In general, a model trained on synthetic data cannot be successfully applied to real data without fine tuning it. But in our case, it becomes possible, probably due to the fact that the simulated data of the coherence decay are quite close to the experimentally observed decay data induced by the environment. 

As final remark, notice that, from the $10^4$ simulated curves $C(\tau,N)$, $6000$ are used for the training of the NNs and $2000$ for their validation. Instead, the test step is performed either by using the remaining $2000$ simulated curves, or by using experimental data as described below.

%%%%%%%%%%%%%%%%%%%%%%%%%%%%%%%
\subsection{Neural networks working principles}
%%%%%%%%%%%%%%%%%%%%%%%%%%%%%%%

Let us describe the main working features of the NNs employed in this paper to carry out noise spectroscopy. Specifically, we are going to use the multi-layer perceptron (MLP) that is composed of fully-connected layers, each of them with a variable number of artificial neurons. 

A single artificial neuron returns as output the scalar 
\begin{equation}
\hat{y} \equiv \Sigma({\bf w}^T\cdot{\bf x}+b)
\end{equation}
that, by definition, is provided by applying the non-linear function $\Sigma:\mathbb{R}\rightarrow\mathbb{R}$ to the weighted sum of the input vector ${\bf x}\in\mathbb{R}^{k}$ to which the bias term $b\in\mathbb{R}$ is added. ${\bf w}\in\mathbb{R}^{k}$ denotes the vector of weights. In our analysis, the activation function $\Sigma$ is chosen equal to the \emph{rectifier} $\Sigma(x) \equiv \max(0,x)$~\cite{glorot2011deep,nair2010rectified}. Thus, a MLP layer composed of $q$ neurons (each with $k$ inputs) returns the vector
\begin{equation}\label{eq:MLP_single_layer}
\hat{{\bf y}} \equiv \Sigma(W^{T}{\bf x}+{\bf b}),
\end{equation}
where $\hat{{\bf y}}\in\mathbb{R}^q$, $W\in\mathbb{R}^{k\times q}$ is the matrix of weights ($W$ collects all the weight vectors of the single neurons), and ${\bf b}\in\mathbb{R}^q$ is the vector of the biases. Hence, a MLP with $L$ layers is ruled by the recursion equation
\begin{equation}\label{eq:MLP_L_layers}
    {\bf h}[\ell] \equiv \Sigma\left(W[\ell]^{T}{\bf h}[\ell-1]+{\bf b}[\ell]\right),
\end{equation}
where $\ell = 1,\ldots,L$ is the index over the number of layers and ${\bf h}[0] \equiv {\bf x}$. In \cref{eq:MLP_L_layers}, $W[\ell]$ and ${\bf b}[\ell]$ are, respectively, the weights and the biases of the $\ell$-th layer. The output vector of the MLP is $\hat{{\bf y}} \equiv {\bf h}[L]$. It is worth noting that the number, dimension and activation functions (they are usually denoted as the \emph{hyperparameters} $\xi$) of the NN layers are chosen through a single optimization routine (cfr.~Methods). 

Let us now introduce the supervised learning process. Ideally, the purpose of the latter is to find the parameters $\theta^{*} = \textnormal{argmin}_{\theta} \mathcal{R}_{\mathcal{D}}(\theta,\xi)$ that minimize the \emph{theoretical risk function}
\begin{equation}\label{eq:risk_function}
    \mathcal{R}_{\mathcal{D}}(\theta,\xi) \equiv \mathbb{E}_{({\bf x},{\bf y})\sim\mathcal{D}}
    \left[\mathcal{L}\left(\hat{{\bf y}},{\bf y}\right)\right],
\end{equation}
where $\theta \equiv \{W[1],{\bf b}[1],\ldots, W[L],{\bf b}[L]\}$, and $\hat{{\bf y}}$ are the estimated values of ${\bf y}$. By definition, $\mathcal{R}_{\mathcal{D}}$ is the expected value of the \emph{loss function} $\mathcal{L}$ for $({\bf x},{\bf y})$ sampled from the distribution $\mathcal{D}$ that generates the dataset~\cite{ShaiUML2014}. The loss function $\mathcal{L}$ is a differentiable function that measures the distance between the prediction $\hat{{\bf y}}$ (output of the MLP) and the desired output ${\bf y}$. However, since one can only dispose of a finite set $\mathcal{S}=\{({\bf x},{\bf y})_1,\dots,({\bf x},{\bf y})_m\}$ of samples to train, validate and test the employed ML models, the theoretical risk function is approximated by the \emph{empirical risk function}. Considering the partition $\{\mathcal{S}_{tr},\mathcal{S}_{va},\mathcal{S}_{te}\}$ of $\mathcal{S}$ in \emph{training} ($\mathcal{S}_{tr}$), \emph{validation} ($\mathcal{S}_{va}$) and \emph{test} ($\mathcal{S}_{te}$) sets, the empirical risk function is defined by:
\begin{equation}
    \mathcal{R}_{\mathcal{S}_{tr}}(\theta,\xi) \equiv \frac{1}{|\mathcal{S}_{tr}|}\sum_{({\bf x},{\bf y})\in \mathcal{S}_{tr}}\mathcal{L}\left(\hat{{\bf y}},{\bf y}\right),
\end{equation}
where $|\mathcal{S}_{tr}|$ is the cardinality of the training set. In fact, $\mathcal{R}_{\mathcal{S}_{tr}}$ is the arithmetic mean of the loss function $\mathcal{L}$ evaluated on the samples of the training set $\mathcal{S}_{tr}$. 

In our paper, we take the loss function $\mathcal{L}$ equal to the \emph{Mean Squared Error} (MSE), also called \emph{L2} loss:  
\begin{equation}
    \mathcal{L}(\hat{\bf y}, {\bf y}) = \frac{1}{q}\sum_{i=1}^{q} (\hat{\bf y}_i - {\bf y}_i)^2
\end{equation}
for the $q$ outputs of the last layer (in our case three, corresponding to the noise parameters $s_0$, $A$, $\sigma$). The MLP is trained by minimizing (step-by-step over time) the empirical risk function $\mathcal{R}_{\mathcal{S}_{tr}}(\theta,\xi)$ with respect to $\theta$ by means of the \emph{mini-batch gradient descent} method, so as to obtain the optimal value $\theta^{*}$ of the NN parameters. Each gradient descent step is defined by
\begin{equation}
    \theta_{t+1} = \theta_t - 
    \eta\nabla_\theta\frac{1}{B}\sum_{b=1}^{B}\mathcal{L}(\hat{\bf y}_{b,t}, {\bf y}_{b,t}),
\end{equation}
where $\theta_0$ is a randomly chosen starting point, $\eta$ is the \emph{learning rate} that defines the length of the step and $\nabla_\theta\frac{1}{B}\sum_{b=1}^B\mathcal{L}(\hat{\bf y}_{t,b}, {\bf y}_{t,b})$ is the gradient of the loss function. The gradient is calculated for any time $t$ on a batch of $B$ elements taken from the training set, and the subscript $\theta$ in $\nabla_\theta$ indicates that the variables of $\mathcal{L}$ during the gradient evaluation are the weights of the NN. In this paper, $\mathcal{R}_{\mathcal{S}_{tr}}$ is minimized by means of Adam~\cite{kingma2014adam} that is a gradient-based optimization algorithm performing the adaptive estimation of lower-order moments. The minimization is stopped when the time-derivative of the risk function evaluated on the validation set $\mathcal{R}_{\mathcal{S}_{va}}(\theta^{*},\xi)$ becomes positive (\emph{early stopping} strategy) or after a predefined number of gradient steps using all the data of the training set (called \emph{epochs}). Then, we use $\mathcal{R}_{\mathcal{S}_{va}}(\theta^{*},\xi)$ to check if the MLP works also for unseen data and tune the hyperparameters $\xi$ (cfr.~Methods). Finally, the test set $\mathcal{S}_{te}$ is employed to calculate the metrics (discussed in detail below) used to generate the figures with the results that we are going to illustrate. 

%%%%%%%%%%%%%%%%%
\subsection{Training and numerical test of neural networks}
\label{subsec:training}
%%%%%%%%%%%%%%%%%

We now show the results obtained by using the trained machine learning models to infer the value of the NSD parameters $\{s_0,A,\sigma\}$. 
As already mentioned, the NNs are tested with $2000$ different NSD parameters. 
For each of these sets of parameters, the curves $C(\tau,N)$ have been simulated as described in the previous subsections. 

In order to determine the smallest amount of data required to reconstruct the NSD, we perform the training, validation and test of the NNs with sub-sets of the simulated curves. These sub-sets are defined by introducing the variable $\overline{N}$ that 
denotes the upper bound for the number of pulses $N\leq\overline{N}$ considered during the whole process. 
For example, for $\overline{N}=16$ only the curves $C(\tau,N)$ with $N\in\{1,8,16\}$ are considered. Note that the sub-sets defined for each value of $\overline{N}$ contain the curves for all the different NSD parameters ($6000$ for training, $2000$ for validation, and $2000$ for testing), and for all the times $\tau$ in the intervals defined in \cref{sec:datasetGeneration}. In detail, the input of the neural network is defined as the concatenation of all the values of $C(\tau,N)$, for $\tau$ in the intervals defined before and $N=1,8,16,...,\overline{N}$. Specifically, ${\bf x}=\{C(\tau_1,1), C(\tau_2,1),\dots,C(\tau_n,1),C(\tau_1,2),\dots,C(\tau_n,2),\dots,C(\tau_n,\overline{N})\}$.

\begin{figure}
    \centering
    \includegraphics[width=0.65\textwidth]{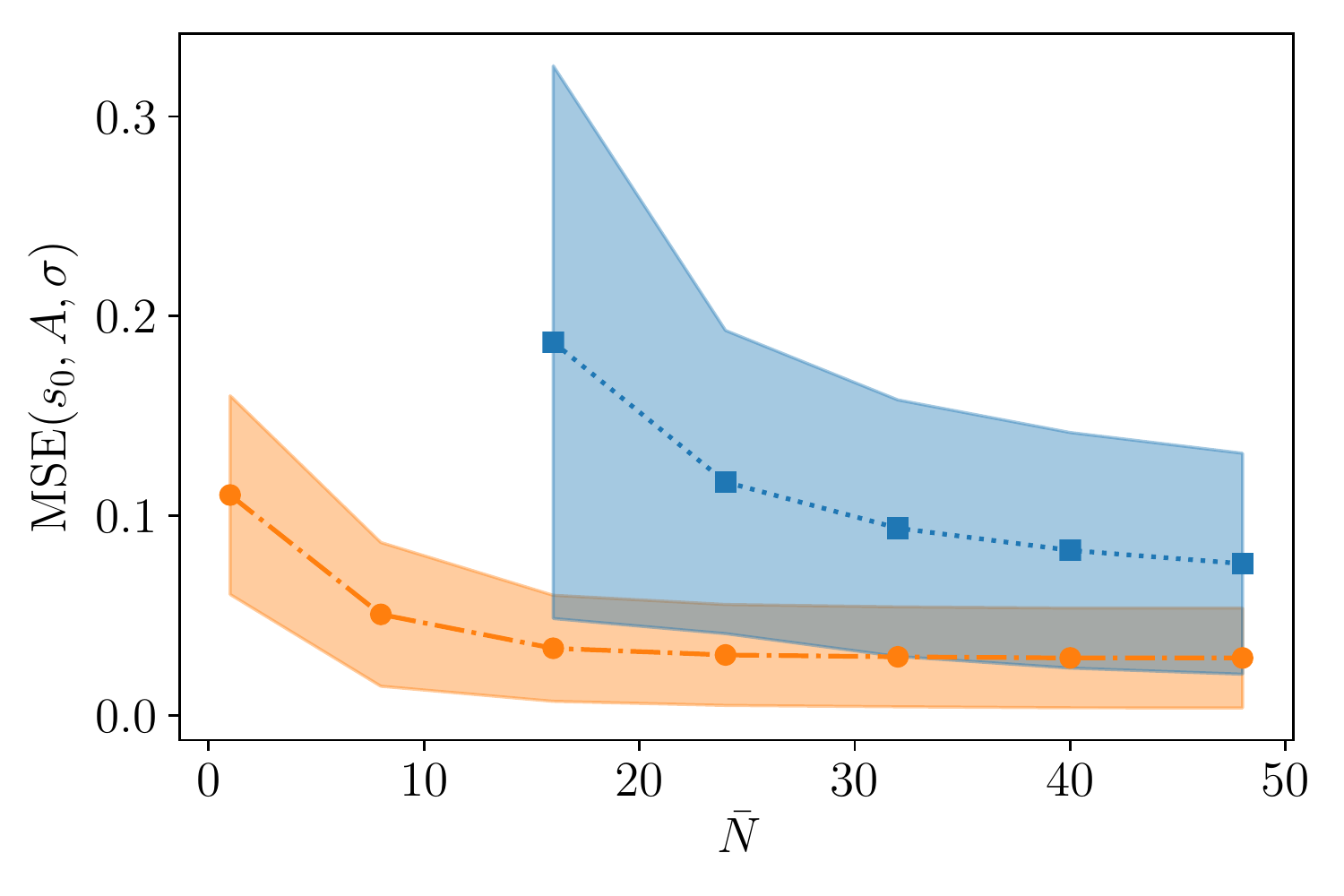}
    \caption{
    Mean-square-errors (MSE) between original and estimated NSD parameters for a set of 2000 test cases. Orange bullets with dash-dotted line are the mean values returned by NNs. Blue squares with dotted line are the mean values provided by the HS method. Finally, shaded areas denote the standard deviation, taking into account all the 2000 cases.
    }
    \label{fig:NSDs_and_MSE_params}
\end{figure}

The results of this analysis are shown in Fig.~\ref{fig:NSDs_and_MSE_params} (orange data), where the MSE (the loss function) between the inferred parameters ($\hat{s_0}$, $\hat{A}$, $\hat{\sigma}$) and the original parameters ($s_0$, $A$, $\sigma$) used to generate the dataset is plotted as a function of $\overline{N}$. 
Remarkably, the MSE seems to achieve its minimum value after $\overline{N}=16$. This entails that the NNs do not significantly improve their precision on the reconstruction of the NSD by using more data to train them beyond this point.

To establish how accurately a NN reconstructs the NSD, we need to compare the corresponding results with those of a different method. 
In particular, we concentrate on the method used in Ref.~\cite{HernandezPRB2018}, which is itself based on Refs.~\cite{AlvarezPRL2011,YugePRL2011}. 
According to them, the decay of the coherence function $C(\tau,N)$ is analyzed as a function of $N$, for each fixed value of $\tau_i$, i.e., for each fixed frequency component of the filter functions. In the limit of high $N$, the decay of the coherence is exponential, with a rate that is inversely proportional to the amplitude of the NSD~\cite{YugePRL2011}. In other words, the amplitude of the NSD is directly estimated for a discrete set of frequencies (each proportional to $1/\tau$). 
In contrast with the original proposals in Refs.~\cite{AlvarezPRL2011,YugePRL2011}, the method in Ref~\cite{HernandezPRB2018} demonstrates that it is better to use the harmonics of the filter functions to reconstruct the NSD, in order to avoid extra broadening of the reconstructed spectrum. For this reason, we denote this method as \emph{Harmonics Spectroscopy} (HS).

We have analyzed the same $2000$ different curves $C(\tau,N)$ (used to test the machine learning models) also with the HS method. The results are collected and shown in Fig.~\ref{fig:NSDs_and_MSE_params} (blue data), where the first point is for $\overline{N}=16$. This is due to the fact that, by definition, the HS method fits the decay of the coherence as a function of $N$. This is possible only for a dataset with at least three points (in this case $N=1,8,16$). As one can observe in Fig.~\ref{fig:NSDs_and_MSE_params},
the MSE values for the HS method (blue region) are always above the MSE values for the NN method (orange region), especially for lower values of $\overline{N}$. These results demonstrate that the NN method can predict the parameters of the NSD with an improved accuracy (up to $5$ times larger) with respect to the HS method. The test presented in this subsection have been performed with simulated data. In the next subsection we are going to repeat the same test but with experimental data.

%%%%%%%%%%%%%%%%%
\subsection{Experimental test of neural networks}
%%%%%%%%%%%%%%%%%

By this point we know that NNs can reliably predict the NSD from \textit{noisy} simulated data. In this section, we want to use the NNs (trained and validated with noisy simulated data) to reconstruct the NSD using experimental data.

\begin{figure}[t!]
    \centering
    \includegraphics[width=\textwidth]{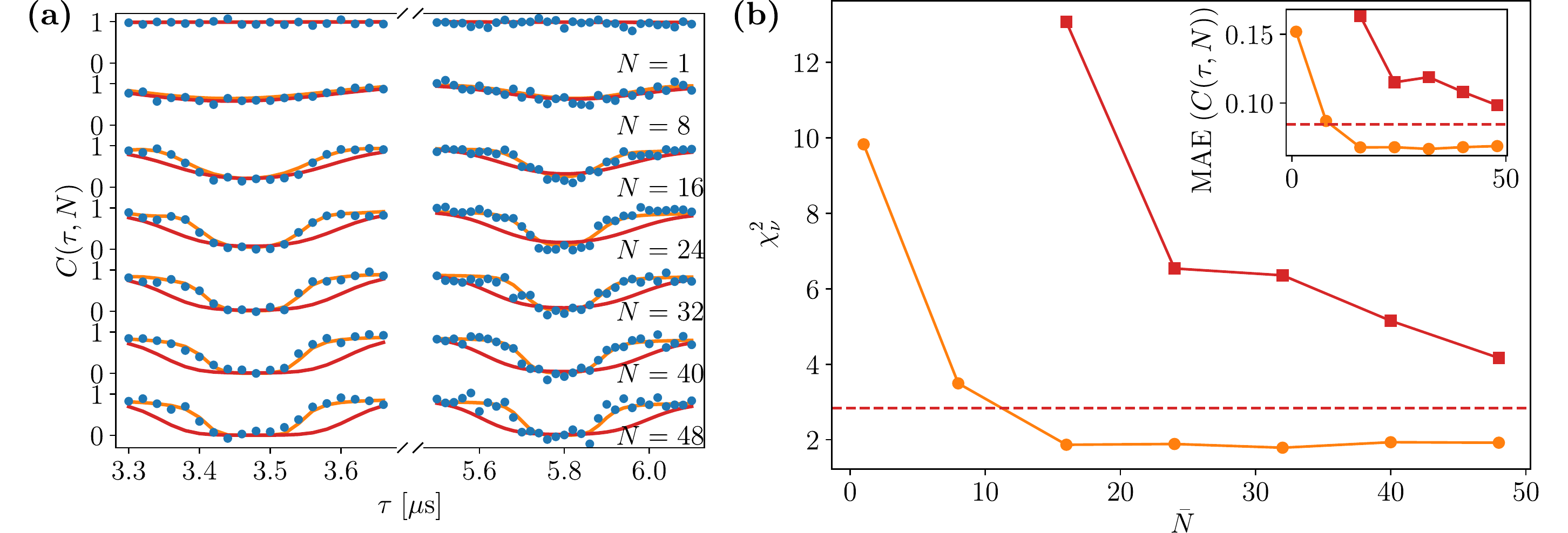}
    \caption{(a) Coherence function $C(\tau,N)$. The experimental data (blue bullets) are shown together with the simulated ones using the NSD predicted respectively by the HS method (red lines) and machine learning models (orange lines), both for $\overline{N}=16$.
    (b) Reduced chi-squared $\chi_\nu^2$, obtained by comparing simulation and experimental data, as a function of $\overline{N}$. As in panel (a), orange and red curves refer to the NN and HS method, respectively. Instead, the dashed line denotes the value of the reduced chi-squared for the HS method when we employ additional measurements for $N=56,64,72,80$ in the interval $\tau \in [5.5 , 6.1]$~$\mu$s.
    Inset: Same results but quantified by the Mean-Absolute-Error (MAE) between the experimental data and the predicted $C(\tau,N)$.}
    \label{fig:experimental_data}
\end{figure}

As quantum sensor we use a spin qubit encoded in the electronic spin of the ground state of a single nitrogen-vacancy (NV) center in a bulk diamond at room temperature. 
This system has proven as a sensitive quantum probe of magnetic fields, with outstanding spacial resolution and sensitivity~\cite{Doherty13,Rondin14}. 
The diamond sample in our experiments has a natural abundance of $^{13}$C impurities ($1.1$\%) that are randomly distributed in the diamond lattice~\cite{TaylorNatPhys2008,GoldsteinPRL2011,AbobeihNatComm2018}.
The $^{13}$C nuclear spins constitute the external environment of the NV center. They act as a collective bath of spins that induces dephasing into the NV electronic spin, limiting the its coherence time $T_2 \approx 100~\mu$s. 
In the presence of strong bias magnetic field ($\geq 150$G)~\cite{ReinhardPRL2012,HernandezPRB2018}, the weak coupling of the NV spin with these carbon impurities can be modeled as a classical stochastic field. The latter has a power spectrum density (here called NSD) that follows a Gaussian distribution centered at the Larmor frequency of the $^{13}$C nuclear spins. 
In order to measure the NV spin coherence function $C(\tau,N)$, we apply a train of $\pi$ pulses (in our case a CP sequence) to the NV spin qubit following the DD protocol described in Fig~\ref{fig:fig1}. 
For more details on the experimental implementation and Hamiltonian of the system see Ref.~\cite{HernandezPRB2018}. 
We have performed this experiment for $N=\{1,8,16,24,32,40,48\}$, and for $\tau \in [3.3 , 3.66]$~$\mu$s and $[5.5 , 6.1]$~$\mu$s with sampling time $\Delta t = 20$ ns. 
The results are shown in Fig.\ref{fig:experimental_data}(a)~(blue bullets). 
Then, the collected coherence functions have been processed and employed to reconstruct the NSD parameters by means of both the NN (trained with the generated dataset) and the HS methods. 
In contrast with the test using simulated data in the previous section, in the experimental case we do not know the exact values of the NSD parameters. Therefore, we cannot calculate the MSE to quantify the accuracy of the reconstructed parameters. In order to estimate such accuracy we have used the following procedure: from the inferred NSD, the coherence curves $C(\tau,N)$ are simulated and then compared with the experimental results. An example of this comparison is shown in Fig.\ref{fig:experimental_data}(a), where $C(\tau,N)$ is simulated under the assumption that the NSD parameters are inferred either by the machine learning models (orange) or by the HS method (red), both for $\overline{N}=16$. Qualitatively it is clear that the orange curves are much closer to the experimental data, than the red curves. 

There are several options to quantitatively compare the experimental data and the simulation results. Here we use both the \emph{reduced chi-squared} $\chi_\nu^2$~\cite{chi}, and the \emph{Mean-Absolute-Error} (MAE)~\cite{mae} between the experimental data and the predicted coherence functions $C(\tau,N)$ (see Methods for more details). The results of this comparison are shown in Fig.~\ref{fig:experimental_data}(b), where $\chi_\nu^2$ and the MAE are plotted as a function of $\overline{N}$. Remarkably, the NSD reconstructed by the NN for $\overline{N}=16$ behaves better that any case using the HS method.
It is worth observing that the same experimental data used to infer the NSD parameters are partially used to estimate the $\chi_\nu^2$ and MAE($C(\tau,N)$). For example, for $\overline{N}=16$, only the data for $N=1,8,16$ are used to reconstruct the NSD, but we employ all the data $N=1,8,16,\dots,48$ to obtain the $\chi_\nu^2$ and MAE($C(\tau,N)$).
Overall, we have observed enhanced performance in reconstructing the NSD of the collective bath of spins, with a maximum improvement (about $7$ times higher) for $\overline{N}=16$. In other words, for $\overline{N}=16$, once we reconstruct the NSD, the quantum sensor dynamics can be predicted with an average square deviation of $\simeq 1.86$ experimental error-bars by using the NN method, or with an average square deviation of $\simeq 13$ error-bars if we use the HS method.

%%%%%%%%%%%%%%%%%%%%
\section{Discussion}
%%%%%%%%%%%%%%%%%%%%

As shown pictorially in Fig.~\ref{fig:fig1}, the NN takes as input the spin qubit coherence functions (the coherence of the quantum sensor decays due to the presence of the external bath) obtained by using a set of different CP control sequences. The NN returns as output the parameters of the unknown NSD in the frequency domain. One can thus note that the NN, once validated, acts as a ``time-frequency converter'' (making use of a quite complicated deconvolution) from the measured signals living in the time domain -- the spin coherence functions -- to the NSD defined in the frequency domain.

The results shown in the previous section, and summarized in Figs.~\ref{fig:NSDs_and_MSE_params} and \ref{fig:experimental_data}(b), demonstrate that NNs can be used to reconstruct the NSD affecting a quantum sensor, achieving higher precision and with considerable less data than the standard HS method. 
Improved values of the reconstruction accuracy have been obtained with simulated and experimental data. 
Both the HS and NN methods are comparable -- in terms of NSD reconstruction accuracy -- for high values of $\overline{N}$, but not for small ones, where NNs give significantly better results. 
Moreover, the main result of our study is that NNs trained with data obtained for $\overline{N}=16$ reconstruct the NSD more accurately than the best estimate provided by the HS method with $\overline{N}=48$.
This improvement is remarkable by itself, but it becomes more significant when we consider that the time required to complete these experiments has a growth faster than a linear function with respect to $\overline{N}$, following an arithmetic progression. As an example, the total time to perform all the experiments in the case of $\overline{N}=16$ and $48$ is respectively $\simeq 10$~minutes and $\simeq 1.2$~hours (for this estimation we consider $10^5$ repetitions as in our experiments, we recall that the total time for each repetition of the single experiment is $T=N\tau$). This is an under-estimation of the time difference between methods, because we are only considering the bare measurement time, without taking into account the time delay between different experiments. Furthermore, it is worth stressing that our results also show that deep learning has a predictive power since it can be applied to never-before-seen data. This naturally provides to the employed machine learning models a connotation of \emph{robustness} that is crucial in real applications.

As a general comment, we stress that it is difficult to identify a definite reason why a machine learning model is more accurate (especially in the case of \textit{small} $\overline{N}$) than a standard DD technique for noise spectroscopy. 
As said above, we observe experimentally that the employed NNs are able to learn non trivial patterns in the sequences of input-output data. 
What the neural network learn is to invert $C(\tau,N)$ as a function of the noise parameters $s_0$, $A$ and $\sigma$ (see Eqs.~(2) and (4) in the main text) that we aim to reconstruct. It is known that NNs are universal approximators of functions: this can be the reason why they are well suited to find the parameters of the NSD from $C(\tau,N)$. 
Moreover, NNs do not consider approximations of the filter function, and they manage to find the noise parameters even for input data containing values of $C(\tau,N)$ with $N$ small. 
In contrast, the \'Alvarez-Suter method, as well as the HS methods in general, arises from approximating the filter function as a Dirac comb. This approximation is not valid for a small number of pulses, hence it is expected to poorly reconstruct the noise spectrum using standard DD techniques. 
In addition, reconstructing the parameters of a NSD using experimental data with a NN trained with synthetic data has been made possible by training the NN over an informative set of noise samples, used to generate the synthetic data. 
The latter, indeed, are given by a collection of values of $C(\tau,N)$ that implicitly include a parameterization of the NSD that is reasonable for the experimental setting; in our case, a Gaussian distribution whose offset, amplitude, width and center belong to finite-valued intervals %containing the corresponding true values with quite high probability.
estimated from similar experimental conditions (e.g. Ref.~\cite{HernandezPRB2018}).

Let us also observe that regression tasks, which are successfully solved by multi-layer perceptrons (one of the easiest form of NN), are less common with respect to the ones to carry out classification; a review of some example datasets and methods for regression is in Ref.~\cite{fernandez2019extensive}. Hence, we expect that the synthetic data used in this work could be useful as a test bed also to the audience of machine learning researchers and developers solving regression problems in different contexts. With this in mind, we share the training dataset with synthetic data and our codes for their generation, as well as the code for machine learning experiments and NSD reconstruction [available on the \emph{GitHub} repository (see Section ``Data and code availability'')]. In this way, we promote the improvement of machine learning models for noise sensing purposes and their use to solve different regression tasks in the quantum estimation framework.

%\section{Conclusions \& outlooks}
\section{Conclusions}

In this paper, we use neural-networks (NNs) to carry out noise spectroscopy with a quantum sensor using dynamical decoupling sequences with a much smaller number of $\pi$ pulses and, at the same time, achieving a higher reconstruction accuracy than standard methods (e.g., HS protocol). This means that with our proposal the noise spectroscopy procedure will take less time and give better results. 
More in detail, we experimentally demonstrate the capability of NNs to reconstruct the NSD of the collective nuclear spin bath that surrounds an electronic spin qubit, i.e., the ground state of a single nitrogen-vacancy center in bulk diamond at room temperature.

To conclude, we outline some possible outlooks for our work. First of all, one may evaluate the performance of NNs that are trained over input data obtained using DD control sequences with more degrees of freedom than the CP ones~\cite{Uhrig07,Zhao11,Souza11,Zhao14,Casanova15}. 
Secondly, deep learning might be applied to noise spectroscopy techniques beyond the HS methods, as for example optimal band-limited control protocols~\cite{FreyNatComm2017,FreyPRApp2020} and even non-Gaussian noise characterization~\cite{PazSilva14,Norris16,Sung19}. In this regard, notice that the NNs take as input the data associated to the spin coherence, and return as output the parameters of the noise spectral density. Therefore, a new NN for the characterization of spin qubit's environment can be trained with coherence curves obtained from using any kind of coherent control sequences. 
The study of the performance of NNs trained with data from these more general control protocols is the next step in understanding how machine learning can enhance quantum sensing. In addition, it might be worth investigating how deep learning can be integrated to quantum sensing procedures that rely on the so-called stochastic quantum Zeno effect~\cite{ShushinJPAMT,GherardiniNJP2016}, whereby the quantum probe is subjected to a sequence of quantum measurements that in the ideal case are designed to confine the dynamics of the probe around the initial (nominal) state~\cite{DoNJP2019,MuellerPLA2020,VirziPRL2022}. We are also confident that the extent of our results can be quite easily replicated in other experimental settings, as e.g., superconducting flux qubits~\cite{Bylander11,Yoshihara14}, trapped ions~\cite{Biercuk09,Kotler11}, cold atoms~\cite{Sagi10,Almog11}, quantum dots~\cite{Chan18,Malinowski17}, NMR experiments in molecules~\cite{AlvarezPRL2011,Fu21}, and nanoelectronic devices~\cite{Muhonen14}. For such a purpose, one might slightly adapt the deep learning techniques used here to methods tailored for time series. 

%%%%%%%%%%%%%%%%%
\section{Methods}
%%%%%%%%%%%%%%%%%

\subsection{Technical details on the training of NNs}

The NN models are developed using the PyTorch framework\,\cite{pytorch} on a machine with 32 CPU cores, 126Gb of RAM and a GeForce RTX 3090 GPU. The training time, including the optimization of the hyperparameters, is around $12$ hours for each $\overline{N}$ .

The hyperparameters optimization is implemented by means of the Ray Tune library\,\cite{liaw2018tune}. The \emph{Hyperopt} package\,\cite{hyperopt} uses the \emph{Tree-structured Parzen Estimators}~\cite{bergstra2011algorithms} algorithm as a Bayesian optimization to search for the best choice of the hyperparameters within a predefined search space. Hyperopt suggest the likely better configurations of the hyperparameters and the underlying model is updated after each trial that is run. The \emph{ASHA} scheduler \cite{asha} is then used to stop the run of the least promising trials chosen by the search algorithm, thus speeding up the hyperparameters optimization process. 

The optimized hyperparameters are the following. \textit{(1)} The \emph{number} of hidden layers decides the value of $L-1$ in \cref{eq:MLP_L_layers}. The hidden layers are between the input layer ${\bf h}[0]$ and the output layer ${\bf h}[L]$. \textit{(2)} The \emph{dimension} of the hidden layers is the value of $q$ in \cref{eq:MLP_single_layer} that, for the sake of simplicity, is equal for all the layers in \cref{eq:MLP_L_layers}. Both the number and dimension of the hidden layers are chosen by sampling log-uniformly an integer value from the space $[1,32)$ and $[1,1024)$, respectively. \textit{(3)} The \emph{learning rate} is responsible for the length of the gradient descent step and it is optimized with a choice between $10^{-2}$, $10^{-3}$ and $10^{-4}$. \textit{(4)} The \emph{batch size} denotes the dimension of the batch on which the loss function is summed for the gradient calculation in a single descent step. The batch size is chosen between $2$, $4$, $8$, $16$, $32$. \textit{(5)} The \emph{dropout} is a regularization strategy that aims to reduce the overfitting by randomly turn off the NN neurons with a predefined probability. Such probability is one among $0$ (no dropout), $0.2$ and $0.5$. \textit{(6)} The \emph{weight decay} is another regularization technique that adds to the loss function the squared weights of the NN multiplied by a decay value. The latter value is optimized choosing between $0$ (no decay), $10^{-6}$, $10^{-5}$, $10^{-4}$ and $10^{-3}$.

To facilitate the reproducibility of the experiments, we summarize in \cref{tab:bestHyperparameters} the optimal values of the hyperparameters for the trained models. Each value of $\overline{N}$ defines the input size of the neural network. Therefore, a different optimization of the hyperparameters is performed for each case.
\begin{table}[h!]
    \centering
    \caption{Hyperparameters for the employed machine learning models. For each value of $\overline{N}$ (that determines the size of the input layer) we report: the number of hidden layers (h.\ l.\ num.), the dimension of each hidden layer (h.\ l.\ dim.) and the values of learning rate (learning r.), batch size, dropout and weight decay (weight d.).}
    \label{tab:bestHyperparameters}
    \begin{tabular}{c|cccccc}
         \textbf{$\overline{N}$} & \textbf{h.\ l.\ num.} & \textbf{h.\ l.\ dim.} & \textbf{learning r.} & \textbf{batch size} & \textbf{dropout} & \textbf{weight d.}   \\
         \hline
         $1$ & $1$ & $2$ & $10^{-2}$ & $16$ & $0$ & $10^{-3}$ \\
         $8$ & $5$ & $328$ & $10^{-4}$ & $4$ & $0$ & $10^{-4}$ \\
         $16$ & $2$ & $133$ & $10^{-3}$ & $8$ & $0$ & $10^{-6}$ \\
         $24$ & $3$ & $224$ & $10^{-4}$ & $2$ & $0$ & $10^{-4}$ \\
         $32$ & $3$ & $145$ & $10^{-4}$ & $4$ & $0$ & $10^{-5}$ \\
         $40$ & $3$ & $286$ & $10^{-4}$ & $4$ & $0$ & $10^{-4}$ \\
         $48$ & $3$ & $38$ & $10^{-3}$ & $8$ & $0$ & $10^{-4}$ \\
    \end{tabular}
\end{table}

\subsection{Definition of quantifiers for reconstruction accuracy}

The accuracy of NN and HS methods can be estimated by using  the reconstructed NSD to simulate the coherence function $C(\tau,N)$, and `measuring' the distance between the simulated data and the experimental values. To do so, we use the \emph{reduced chi-squared} $\chi_\nu^2$, and the Mean-Absolute-Error (MAE($C$)): 
We define $C_{\rm e}\pm \delta C_{\rm e}$~($C_{\rm s}$) as the experimental~(simulated) values of $C(\tau,N)$, where $\delta C_{\rm e}$ is the standard deviation of the experimental data. 
Then we can write \emph{reduced chi-squared} and the MAE as
\begin{eqnarray}
    \chi_\nu^2 &\equiv& \frac{1}{\nu}\sum_{n,N} \frac{(C_{\rm e}(\tau_n,N) -C_{\rm s}(\tau_n,N) )^2}{ \delta C_{\rm e}(\tau_n,N)^2} \\
    {\rm MAE}(C) &\equiv& \frac{1}{\nu}\sum_{n,N} \left|C_{\rm e}(\tau_n,N) -C_{\rm s}(\tau_n,N) \right| , 
\end{eqnarray}
where $N=\{1,8,16,24,\dots,\overline{N}\}$, $\{\tau_n\}$ are the values of the time between pulses within the time intervals defined in main text, and $\nu$ is the total number of elements in the sum. 
Notice that $\chi_\nu^2$ takes into account the experimental precision to scale the difference between experiment and simulation.
The results showing both $\chi_\nu^2$ and the MAE are in Fig.~\ref{fig:experimental_data}.

%%%%%%%%%%%%%%%%%%%%%%%%%%%%%%%%%%%%
\section*{Data and code availability}
%%%%%%%%%%%%%%%%%%%%%%%%%%%%%%%%%%%%

The source codes for the generation of the training dataset and the machine learning experiments are available on \emph{GitHub} at the following address:
\href{https://github.com/trianam/noiseSpectroscopyNV}{https://github.com/trianam/noiseSpectroscopyNV}

%%%%%%%%%%%%%%%%%%%%%%
%\section{Acknowledgements}
%%%%%%%%%%%%%%%%%%%%%%
\section*{Acknowledgements}
This work was supported by the European Commission’s Horizon Europe Framework Programme under the Research and Innovation Action GA n.\,101070546–MUQUABIS, and by the European Defence Agency under the project Q-LAMPS Contract No B PRJ-RT-989. S. H. G. acknowledges support from CNR-FOE-LENS-2020. F. C. and S. M. acknowledge the European Union’s Horizon 2020 research and innovation programme under FET-OPEN GA n.\,828946--PATHOS.

\section*{References}
\bibliographystyle{unsrt}
\bibliography{biblio}

\end{document}